\documentstyle[preprint,aps]{revtex}

\draft

\newcommand{\be}{\begin{equation}}
\newcommand{\ee}{\end{equation}}
\newcommand{\ben}{\begin{eqnarray}}
\newcommand{\een}{\end{eqnarray}}

\newcommand{\iii}{\'{\i}}

\begin{document}

\draft
\title{Understanding Quantum Entanglement: Qubits, Rebits and the Quaternionic
Approach}
\author{J. Batle$^1$\footnote{E-mail:
vdfsjbv4@uib.es, corresponding author.}, A. R. Plastino$^{1,\,2,\,3}$,
M. Casas$^1$, and A. Plastino$^{2,\,3}$}

\address {$^1$Departament de F\iii sica, Universitat de les Illes Balears,
07071 Palma de Mallorca, Spain \\
$^2$National University La Plata, C.C. 727, 1900 La Plata, Argentina \\
  $^3$Argentina's National Research Council (CONICET) }

\date{\today}

\maketitle
 \begin{abstract}

 It has been recently pointed out by Caves, Fuchs, and Rungta \cite{CFR01}
 that real quantum mechanics (that is, quantum
mechanics defined over real vector spaces \cite{S60,GPRS61,E86,W02}) provides
an interesting foil theory whose study may shed some light on just which
particular aspects of quantum entanglement are unique to standard quantum
theory, and which ones are more generic over other physical theories endowed
with this
phenomenon. Following this work, we discuss 
some entanglement properties of two-rebits systems, making a comparison with
the basic properties of two-qubits systems, i.e., the ones described by
standard complex quantum mechanics. We also discuss the  use of quaternionic
quantum mechanics as applied to the phenomenon of entanglement.

\vskip 5mm
 Pacs: 03.67.-a; 89.70.+c; 03.65.-w

\vskip 5mm

\noindent  Keywords: Quantum Entanglement; Quantum Information
Theory; Real Quantum Mechanics; Quaternionic Quantum Mechanics

\end{abstract}
\vspace{.5cm}

\maketitle

\newpage

\section{Introduction}

The phenomenon of entanglement
is one of the most characteristic non-classical features
exhibited by quantum systems \cite{LPS98}. Quantum entanglement is
the basic resource of several quantum processes as superdense coding
\cite{BW93}, quantum teleportation \cite{BBCJPW93}
and quantum computation \cite{LPS98,GD02}
studied by quantum information theory
\cite{WC97,W98,BEZ00,AB01}.
A state of a composite quantum system constituted by subsystems
$A$ and $B$ is called ``entangled" if it cannot be represented as
a mixture of factorizable pure states.

The simplest systems exhibiting the phenomenon of entanglement are two-qubits
systems. For these systems an explicit expression of the entanglement of
formation, which is a natural quantitative measure of entanglement
\cite{BDSW96}, has been found by Wootters \cite{WO98}. The correspondent space
of mixed two-qubits states in standard quantum mechanics is 15-dimensional. The
amount of  entanglement in this space has been established by Wootters
\cite{ZHS98,Z99,MJWK01,IH00}.

  For quantum mechanics defined over real vector spaces the
  simplest composite systems are two-rebits systems.
  An explicit expression for the
  entanglement of formation of arbitrary states of
  two-rebits has been obtained by Caves, Fuchs and
  Rungta \cite{CFR01}. Pure states of rebits-systems are described by
  normalized vectors in a two dimensional real vector space.
 The correspondent space of mixed two-rebits states is 9-dimensional (vis-\`a-vis 15 for 2 qubits).

 The aim of the present work is to explore numerically, as well
 as conceptually, the entanglement properties of two-rebits systems \cite{JB02},
 as compared to the usual two-qubits ones, so as to detect  the
 differences between the two types of system. We pay particular attention to the distribution
of states according to their degree of entanglement. We obtain, analytically
the
 probability densities for finding pure states
states with a given amount of entanglement $E$ (or with fixed values of the
 so called concurrence squared $C^2$). The same is done for mixed states in
 numerical fashion.

  The paper is organized as follows. In section II we review the general
 properties of two-qubits systems. Several quantities
 related to the entanglement of formation are investigated for both rebits and
 qubits in section III, with an emphasis on the differences and similitudes between  both
 formalisms.
  In section IV we discuss how to follow a quaternionic
 approach to quantum entanglement. Finally, some conclusions
  are drawn in section V.

\section{Entanglement of two-qubits systems}

As already stressed in the introduction, the two-qubits system is the simplest
quantum mechanical system that exhibits the sort of ``extra correlation" one
calls entanglement between  two parties. A representation of a qubit is given
by the Bloch sphere. The poles correspond to classical bits $|0\rangle$,
$|1\rangle$ while any point on this unit sphere, given by a pair of angles
$\phi,\psi$, represents a general qubit cos$\phi |0\rangle+e^{i\psi}$sin$\phi
|1\rangle$. Qubits constitute the essential new elements in the switch from
Classical to Quantum Information Theory and entanglement the basic ingredient
of  many  striking processes now under active investigation.

Here we use the concurrence $C[\hat \rho ]$ and the entanglement of formation
 $E[\hat \rho]$  as  quantitative measures of the amount of entanglement.
  Wootters' formula for the entanglement of formation of a
 two-qubits state $\hat \rho$ reads \cite{WO98}

\begin{equation} \label{enta}
E[\hat \rho] \, = \, h\left( \frac{1+\sqrt{1-C^2}}{2}\right), \end{equation}

\noindent where

\begin{equation}\label{enta1}
h(x) \, = \, -x \log_2 x \, - \, (1-x)\log_2(1-x), \end{equation}

\noindent and the concurrence $C$ is given by $C \, = \,
max(0,\lambda_1-\lambda_2-\lambda_3-\lambda_4)$, with $\lambda_i, \,\,\, (i=1,
\ldots, 4)$ being the square roots, in decreasing order, of the eigenvalues of
the matrix $\hat \rho \tilde \rho$, with

\begin{equation} \label{rhotil} \tilde \rho \, = \, (\sigma_y \otimes \sigma_y)
\rho^{*} (\sigma_y \otimes \sigma_y). \end{equation}

\noindent The above expression has to be evaluated by recourse to the matrix
elements of $\hat \rho$ computed  with respect to the conventional product
basis.

\vskip 4mm
  In this work we have generated {\it all} pure and mixed states of a two-qubits system according
to the measure defined in \cite{ZHS98,Z99,PZK98}.

  The distribution of two-qubit states according
  to their degree of mixture can be obtained analytically adopting a
  geometric representation \cite{nostro} for the eigenvalues of $\hat \rho$
 as a regular tetrahedron $T_{\Delta}$
  of side length 1, in ${\cal R}^3$, centered at the origin.
  There exists a mapping connecting these eigenvalues $(p_1,\ldots, p_4)$
  with the points $\bf r$ of the tetrahedron, so that we can relate
  the participation ratio $R(\hat \rho)=1/Tr(\hat \rho^2)$ to the radius
        $r=\mid {\bf r} \mid$
  of a sphere embedded within the tetrahedron $T_{\Delta}$ according to

  \begin{equation} \label{tetra2}
  r^2 \, = \, -\frac{1}{8} \, + \, \frac{1}{2} \sum_{i=1}^4 p_i^2.
  \end{equation}

 \noindent
 Thus, the states with a given degree of mixture lie on the
 surface of a sphere of radius $r$ concentric with the
 tetrahedron $T_{\Delta}$. For a completely description, see \cite{nostro}. It
 is then easy to find out just how our states are so distributed, according to the
degree of mixture $R$. This distribution is depicted in Figure 1. The dualistic
nature of the concepts of  entanglement and mixedness becomes apparent. For
two-qubits
 systems one would expect that
 the states will tend to become less entangled as the degree of mixture increases.
 In fact, for $R \ge 3$
(vertical line in Figure 1) {\it all states are separable}. It is important to
stress here that this curve is
 the same no matter what quantities (rebits, qubits, or even quaterbits) are involved,
although their respective entanglement properties are different.

\section{Entanglement properties of two-rebits systems. A comparison}

In the space of real quantum mechanics we can represent  rebits on the Bloch
sphere. The poles  correspond to classical bits $|0\rangle$, $|1\rangle$, but
the sphere  reduces itself now to a maximum unit circle, described by just one
parameter $\phi$. We have cos$\phi |0\rangle+e^{i\psi}$sin$\phi |1\rangle$
$\rightarrow$ cos$\phi |0\rangle+$sin$\phi |1\rangle$. Entanglement can also be
described in such a context with suitable modifications. Caves, Fuchs, and
Rungta's (CFR) formula  for the entanglement of formation of a two-rebits state
$\rho$ is given by (\ref{enta}), (\ref{enta1}) using the concurrence 
$C[\rho] \, = \, \mid \!
{\rm tr} (\tau) \! \mid \, = \,
 \mid \! {\rm tr} (\rho \, \sigma_y \otimes \sigma_y) \! \mid$, which has to
be evaluated using the matrix elements of $\rho$ computed  with respect to the
product basis, $\mid \! i,j \rangle = \mid i \rangle \! \mid j
\rangle, \,\, i,j=0,1$.

  For a two-rebits state the entanglement of formation is completely determined
  by the expectation value of one single observable,
  namely, $\sigma_y \otimes \sigma_y$.
  On the contrary, it has been recently proved that there
  is no observable (not even for pure states) whose sole
  expectation value will yield enough information to as to determine
  the entanglement of a two-qubits state \cite{SH00}.
As  shown in \cite{JB02}, there are mixed
  states of two rebits with maximum entanglement
  (that is, with $C^2=1$) within the range $1\le R \le 2$.
        This is clearly in contrast to what happens with two-qubits states,
 because only pure states ($R=1$) have maximum entanglement.

 The  measure of mixedness $R$ introduced above
can be used in the description of the entanglement phenomenon for two-rebits
systems. Also, in order to explore numerically the properties of arbitrary
two-rebits states, it is necessary to introduce an appropriate measure $\mu $
on the space ${\cal S}_R$ of general two-rebits states. Such a measure is
needed to compute volumes within the space ${\cal S}_R$, as well as to
determine what is to be understood by a uniform distribution of states on
${\cal S}_R$  \cite{ZHS98,Z99}.

 An arbitrary (pure and mixed) state $\rho$ of
a (real) quantum system described by an $N$-dimensional real
Hilbert space can always be expressed as the product of three
matrices,

\begin{equation}\label{odot} \rho \, = \, R D[\{\lambda_i\}] R^{T}. \end{equation}

\noindent Here $R$ is an $N\times N$ orthogonal matrix and
$D[\{\lambda_i\}]$ is an $N\times N$ diagonal  matrix whose
diagonal elements are $\{\lambda_1, \ldots, \lambda_N \}$, with $0
\le \lambda_i \le 1$, and $\sum_i \lambda_i = 1$.
   The group of orthogonal matrices $O(N)$ is
endowed with a unique, uniform measure $\nu$ \cite{PZK98}. On the
other hand, the simplex $\Delta$, consisting of all the real
$N$-uples $\{\lambda_1, \ldots, \lambda_N \}$ appearing in
(\ref{odot}), is a subset of a $(N-1)$-dimensional hyperplane of
${\cal R}^N$. Consequently, the standard normalized Lebesgue
measure ${\cal L}_{N-1}$ on ${\cal R}^{N-1}$ provides a natural
measure for $\Delta$. The aforementioned measures on $O(N)$ and
$\Delta$ lead then to a natural measure $\mu = \nu \times {\cal L}_{N-1}$
on the set ${\cal S}_R$ of all the states of our (real) quantum system.

 Clearly, our system will have $N=4$ again, and in our
 numerical computations we are going to randomly generate
 states of a two-rebits system according to the measure $\mu$.

The relationship between the amount of entanglement and the purity
of quantum states of composite systems has been recently discussed
in the literature \cite{ZHS98,Z99,MJWK01,IH00}.
  The amount of entanglement and the purity of quantum states of composite
  systems exhibit a dualistic relationship. As the degree of
  mixture increases, quantum states tend to have a smaller
  amount of entanglement. In the case of two-qubits systems,
  states with a large enough degree of mixture are always
  separable \cite{ZHS98}.
  To study the relationship between entanglement and mixture in real
quantum mechanics,
we compute numerically the probability $P(E)$ of finding
 a two-rebits state endowed with an amount of entanglement $E$.
In Figure 2 we compare (i) the distribution associated with two-rebits states
with (ii) the one, associated with two-qubits states,  recently obtained by
Zyczkowski {\it et al.} \cite{ZHS98,ZS01}. Fig. 2a depicts the probability
$P(E)$ of finding two-qubits states endowed with a given entanglement $E$ (as
computed with Wootters' expression). In a similar way, Fig. 2b exhibits a plot
of the probability $P(E)$ of finding two-rebits states endowed with a given
entanglement $E$ (as computed with the CFR formula). Comparing Figs 2a and 2b
we find that the distributions $P(E)$ describing arbitrary states (that is,
both pure and mixed states) exhibit the same qualitative shape for both
two-qubits and two-rebits states: in the two cases the distribution $P(E)$ is a
decreasing function of $E$.

  The distribution $P(E)$ or $P(C^2)$ for pure two-rebits
  states can be
  obtained analytically. Let us write a pure two-rebits state
  in the form

 \begin{equation}\label{repuro}
\mid \Psi \rangle \, = \,  \sum_{i=1}^4 \, c_i
  \mid \phi_{i}\rangle,
 \end{equation}

 \noindent
 where

 \begin{equation}\label{esfera4}
 \sum_{i=1}^4 \, c_i^2 \, = \, 1, \,\,\,\,\,\, c_i\in {\cal R}.
 \end{equation}

 \noindent
 The states $( \mid \phi_{i}\rangle, \,\,\, i=1,\ldots, 4$)
 are the eigenstates of the operator $\sigma_y \otimes \sigma_y$. The four real
 numbers $c_i$ constitute the coordinates of a point lying on
 the three dimensional unitary hyper-sphere $S_3$ (which is
 embedded in ${\cal R}^4$). We now introduce on $S_3$ three
 angular coordinates, $\phi_{1}$, $\phi_{2}$, and $\theta $,
 defined by

 \ben \label{triangle}
 c_1 \, &=& \, \cos \theta \cos \phi_1, \cr
 c_2 \, &=& \, \cos \theta \sin \phi_1, \cr
 c_3 \, &=& \, \sin \theta \cos \phi_2, \cr
 c_4 \, &=& \, \sin \theta \sin \phi_2, \,\,\,\,\,0\le \theta < \frac{\pi}{2},
 \,\, 0\le \phi_1,\phi_2 <2\pi.
 \een

\noindent In terms of the above angular coordinates, the
concurrence of the pure state $\mid \!\Psi \rangle $ is given by

\begin{equation} \label{cetita} C \, = \, \mid \! \langle \sigma_y \otimes
\sigma_y \rangle \! \mid \, = \, \mid \!\cos 2\theta \!\mid. \end{equation}

Using (\ref{triangle}) and (\ref{cetita}) one deduces that the probability
density $P(C^2)$ of finding a pure two-rebits state with a  squared concurrence
$C^2$ is given by

\begin{equation}
 P(C^2) \, = \, \frac{1}{2\sqrt{C^2}}.
\end{equation}

\noindent The distribution is to be compared with the one obtained for pure
states of two-qubits systems, which is (analytically) found to be \cite{ZS01}

\begin{equation}
 P(C^2) \, = \, \frac{3}{2}\sqrt{1-C^2}.
\end{equation}

Both distributions are compared in Figure 3. Figure 3a depicts the one for
qubits, while Fig. 3b shows the one for rebits. The distribution remains
finite, in the case of  qubits, for all $C^2$. In the case of rebits it
presents a sharp peak at the origin, an then saturates to 1/2 at $C^2=1$. The
general conclusion that one draws from Figures 2 and 3 is that the curves
representing the distributions $P(E)$ and $P(C^2)$ associated with (i) pure
states and (ii) arbitrary states do not differ, in the case of
two-rebits states,  as much as they do in the case of two-qubits states.

 We can determine analytically which is the maximum entanglement
 $E_m$ of a two-rebits state compatible with a given participation ratio $R$.
 Since $E$ is a monotonic increasing function of the
 concurrence $C$, we shall find the maximum value of $C$
 compatible with a given value of $R$. In order to solve the
 ensuing variational problem (and bearing in mind that
 $C = \mid \!\! \langle \sigma_y \otimes \sigma_y \rangle \!\! \mid$ ),
 let us {\it first} find the state that extremizes
  ${\rm Tr} (\rho^2)$ under the constraints associated with a
  given value of  $\langle \sigma_y \otimes \sigma_y \rangle $,
 and the normalization of $\rho $. This variational
 problem can be cast in the fashion

\begin{equation} \label{maxent1}
 \delta \Bigl[ {\rm Tr} (\rho^2) \, + \, \beta
\langle \sigma_y \otimes \sigma_y \rangle -\alpha {\rm Tr} (\rho)
\Bigr] \, = \, 0, \end{equation}

\noindent where $\alpha $ and $\beta $ are appropriate Lagrange
multipliers.

 After some algebra, and  expressing  the expectation value of
$\langle \sigma_y \otimes \sigma_y \rangle $ in terms of the parameter $\beta$,
 one finds that the maximum value of $C^2$ compatible with a given value
 of $R$ is given by

 \begin{equation} \label{cedoser}
 C^2_m \, = \, \left\{ 1 \,\,\,\,\,\,\,\,\,\,\,\,\,\, ; \,\,\, 1\le R \le 2
 \atop \frac{4}{R}-1 \,\,\,         ; \,\,\, 2\le R \le 4. \right.
 \end{equation}

\section{The quaternionic approach to quantum entanglement}

The quaternionic space $\cal{H}$ constitutes a generalization of the complex
space $\cal{C}$ which, in turn, generalizes the real space $\cal{R}$.
Each step of this chain is
possible by introducing new quantities: $i^2=-1$ from $\cal{R}$ to $\cal{C}$ and
$j^2=k^2=-1$ from $\cal{C}$ to $\cal{H}$, with suitable commutation laws for the
three quantities $i,j,k$.

A general quaternion $\phi$ and its associated commutation algebra are written
in the following fashion

 \begin{equation} \label{quater}
 \phi = \phi_0 + i\phi_1 + j\phi_2 + k\phi_3,
 \end{equation}

\noindent with $\phi_i \in \cal{R}$ and
 $ij = -ji = k$, $jk = -kj = i$, $ki = -ik = j$.

This ``natural extension" of complex numbers that yields quaternions cannot be
generalized any  further. Thus, if we give up the
property of commutativity, the most general algebra that can be used in quantum
mechanics is the quaternionic one \cite{A95}.

\subsection{Entanglement for pure states of two ``quaterbits" systems}

The definition of entanglement in Quaternionic Quantum Mechanics (QQM) for pure
states does not differ from the standard one. Given a pure state $|\psi\rangle$
of a composite bipartite system, the entanglement is obtained via the von
Neumann entropy of the marginal density matrix associated to subsystem A by
tracing over the subsystem B: $\hat \rho_A=$Tr$_B |\psi\rangle_{AB} \langle
\psi|$ of the total density matrix $\hat \rho = |\psi\rangle_{AB} \langle
\psi|$, or, vice versa, $\hat \rho_B=$Tr$_A |\psi\rangle_{AB} \langle \psi|$.
Thus, $E(\hat \rho)=S(\hat \rho_A)=S(\hat \rho_B)$.

In the case of quaternions we face a higher dimensionality and, therefore, we
need more parameters to describe the state $\hat \rho$. Additionally, in using
the  kets and bras notation of Dirac's we must keep in mind  the quaternions'
non-commutativity rules. For the sake of simplicity, let us suppose that a pure
state is written as

\begin{equation}
|\psi\rangle = C_1 |0\rangle_{A}|0\rangle_{B} + C_2 |1\rangle_{A}|1\rangle_{B};\
C_1, C_2 \in {\cal{H}},\ |C_1|^2+|C_2|^2=1.
\end{equation}

 The statistical matrix $\hat \rho=|\psi\rangle_{AB}\langle \psi|$ reads

\begin{equation}
\left( \begin{array}{c} C_1 \\ 0 \\ 0 \\ C_2 \end{array} \right)
\left(
\begin{array}{cccc} \overline{C_1} & 0 & 0 & \overline{C_2}\end{array}
\right) = \left( \begin{array}{cccc}
C_1 \overline{C_1} & 0 & 0 & C_1 \overline{C_2}\\
0 & 0 & 0 & 0\\
0 & 0 & 0 & 0\\
C_2 \overline{C_1} & 0 & 0 & C_2 \overline{C_2} \end{array} \right),
\end{equation}

\noindent with $\hat \rho^{\dag}=\hat \rho$  since $\overline{C_1
\overline{C_2}}=\overline{\overline{C_2}} \ \overline{C_1}= C_2
\overline{C_1}$. Entanglement then is only a function of
$N(C_1)^2=C_{10}^2+C_{11}^2+C_{12}^2+C_{13}^2$. Identifying $ x\equiv
C_{10}^2+C_{11}^2$ and $y\equiv C_{12}^2+C_{13}^2$, we plot $E(\hat\rho)$ in
Fig. 4, together with $E(\hat\rho)$ for the three  Quantum Mechanics' versions
we are dealing with here.

\subsection{Extension to general mixed states}

The full analytical study of entanglement in the framework of Quaternionic
Quantum Mechanics requires careful consideration  of the algebra of states and
operators for these ``hyper-numbers". If one wishes to discuss how to carry out 
statistical studies and  how entanglement-related properties are distributed
over the space of all (pure and mixed) states, one notices that in this case the
dimensionality of the problem for a general 4 x 4 matrix is substantially
higher (3+$4*$6=27) than for the complex (3+$2*$6=15) or the real (3+$1*$6=9)
cases. The ensuing statistical properties become clearly non-trivial and some 
substantial effort is required.

Let us merely list here the basic ingredients needed for a complete description
of the statistical properties of quaternionic states $\hat\rho$: i) to build
and correctly parametrize the unitary transformations defined over the
quaternionic Hilbert space. ii) to specify  the form of the Haar measure for
the concomitant space of unitary transformations. Notice that the measure on
the simplex is exactly the same as for complex or real systems. For mixed
states $\hat \rho$, the distributions associated to quantities that depend only
on the eigenvalues of a statistical operator do have the same form in the real
or the complex cases, for they only depend on the simplex.  The results of the
corresponding numerical study will be published elsewhere.


\section{Conclusions}

We have explored numerically the entanglement properties of two-rebits systems.
A systematic comparison has been established between many statistical
properties of two-qubits and two-rebits systems. We paid particular attention
to the relationship between entanglement and purity in both quantum mechanical
frameworks. We have also determined numerically the probability densities
$P(E)$ of finding (i) pure two-rebits states and (ii) arbitrary two-rebits
states, endowed with a given amount of entanglement $E$ or concurrence squared
$C^2$. In particular, we determined analytically the maximum possible value of
the concurrence squared $C^2$ of two-rebits states compatible with a given
value of mixedness $R$. As for the probability of finding states with a given
amount of entanglement, the difference between mixed and pure sates is much
larger for qubits than for rebits. Also, we have sketched the manner in which
the quaternionic formalism could be applied to the study of quantum
entanglement.

\acknowledgments This work was partially supported by the  DGES grants
PB98-0124  (Spain), and by CONICET (Argentine Agency).

\newpage

\noindent {\bf FIGURE CAPTIONS}

\vskip 0.5cm

\noindent Fig. 1- Probability (density) distribution for finding a
state $\hat \rho$ with a given participation ratio R. States (two-qubits)
 with $R \geq 3$ are always separable.

\vskip 0.5cm

\noindent  Fig. 2- Probability (density) distributions for finding a
state $\hat \rho$ with a given amount of entanglement $E$. Fig. a corresponds to
qubits, while Fig. b to rebits.

\vskip 0.5cm

\noindent Fig. 3- Probability distributions for finding a
state $\hat \rho$ with a given concurrence squared $C^2$. Fig. a corresponds to
qubits, while Fig. b to rebits.

\vskip 0.5cm

\noindent  Fig. 4- Plot of the entanglement of pure states in the framework of
Real Quantum Mechanics (RQM), Standard Quantum Mechanics (SQM) and Quaternionic
Quantum Mechanics (QQM). See text for details.

\end{document}